# Emergence of the persistent spin helix in semiconductor quantum wells

Jake D. Koralek[1], Chris Weber[1], Joe Orenstein[1,2], Andrei Bernevig[3], Shoucheng Zhang[4], Shawn Mack[5], David Awschalom[5]

[1]Materials Science Division, Lawrence Berkeley National Laboratory, Berkeley, California 94720, USA

[2]Department of Physics, University of California Berkeley, Berkeley, California 94720, USA

[3] Princeton Center for Theoretical Science, Princeton University, Princeton, New Jersey 08540, USA

[4]Department of Physics, Stanford University, Stanford, California, 94305, USA

[5]Center for Spintronics and Quantum Computation, University of California, Santa Barbara, California, 93106, USA

According to Noether's theorem,[1] for every symmetry in nature there is a corresponding conservation law. For example, invariance with respect to spatial translation corresponds to conservation of momentum. In another well-known example, invariance with respect to rotation of the electron's spin, or SU(2) symmetry, leads to conservation of spin polarization. For electrons in a solid, this symmetry is ordinarily broken by spin-orbit (SO) coupling, allowing spin angular momentum to flow to orbital angular momentum. However, it has recently been predicted that SU(2) can be recovered in a two-dimensional electron gas (2DEG), despite the presence of SO coupling.[2] The corresponding conserved quantities include the amplitude and phase of a helical spin density wave termed the "persistent spin helix" (PSH).[2] SU(2) is restored, in principle, when the strength of two dominant SO interactions, the Rashba[3] ($\alpha$) and linear Dresselhaus[4] ($\beta_1$), are equal. This symmetry is predicted to be robust against all forms of spin-independent scattering, including electron-electron interactions, but is broken by the cubic Dresselhaus term ($\beta_3$) and spin-dependent scattering. When these terms are negligible, the distance over which spin information can propagate is predicted to diverge as $\alpha \to \beta_1$. Here we observe experimentally the emergence of the PSH in GaAs quantum wells (QW's) by independently tuning $\alpha$ and $\beta_1$. Using transient spin-grating spectroscopy[5] (TSG), we find a spin-lifetime enhancement of two orders of magnitude near the symmetry point. Excellent quantitative agreement with theory across a wide range of sample parameters allows us to obtain an absolute measure of all relevant SO terms, identifying $\beta_3$ as the main SU(2) violating term in our samples. The tunable suppression of spin-relaxation demonstrated in this work is well-suited for application to spintronics.[6,7]

TSG spectroscopy is a powerful tool to search for the PSH because it enables one to measure the lifetime of spin polarization waves as a function of wavevector. In TSG, spin polarization waves of well-defined **q** are generated by exciting a 2DEG with two non-colinear beams of light from a femtosecond laser. When the two incident pulses of light are linearly polarized in orthogonal directions, interference generates stripes of alternating photon helicity in the sample. Because of the optical orientation[8] effect in III-V semiconductors, the photon helicity wave generates a spin polarization wave in the 2DEG. The wavevector is varied by changing the angle between the interfering beams. The spin wave imprinted in the 2DEG acts as an optical diffraction grating, allowing its subsequent temporal evolution to be monitored by the diffraction of a time-delayed probe pulse.[9]

In Fig. 1a we show a set of TSG decay curves for a 2DEG residing in an asymmetrically modulation doped GaAs QW, which is expected to have both Rashba and Dresselhaus SO interactions. Each curve represents the decay of a spin grating at a specific **q**. The decay at $q = 0$ (not shown), measured by time-resolved Faraday rotation,[10] follows a single exponential over nearly 3 orders of magnitude. With increasing $q$ the decay evolves towards the sum of two exponentials with nearly equal weight, but very different rate constants. We fit the TSG decay curves with double exponentials and plot the resulting spin-lifetimes in figure 1b, immediately revealing very unusual spin-diffusion properties. The rapidly decaying component of the TSG (labeled $\tau_R$) displays ordinary diffusion in the sense that the spin-lifetime is peaked at q=0. The slowly decaying component ($\tau_E$), on the other hand, is peaked sharply at non-zero q.

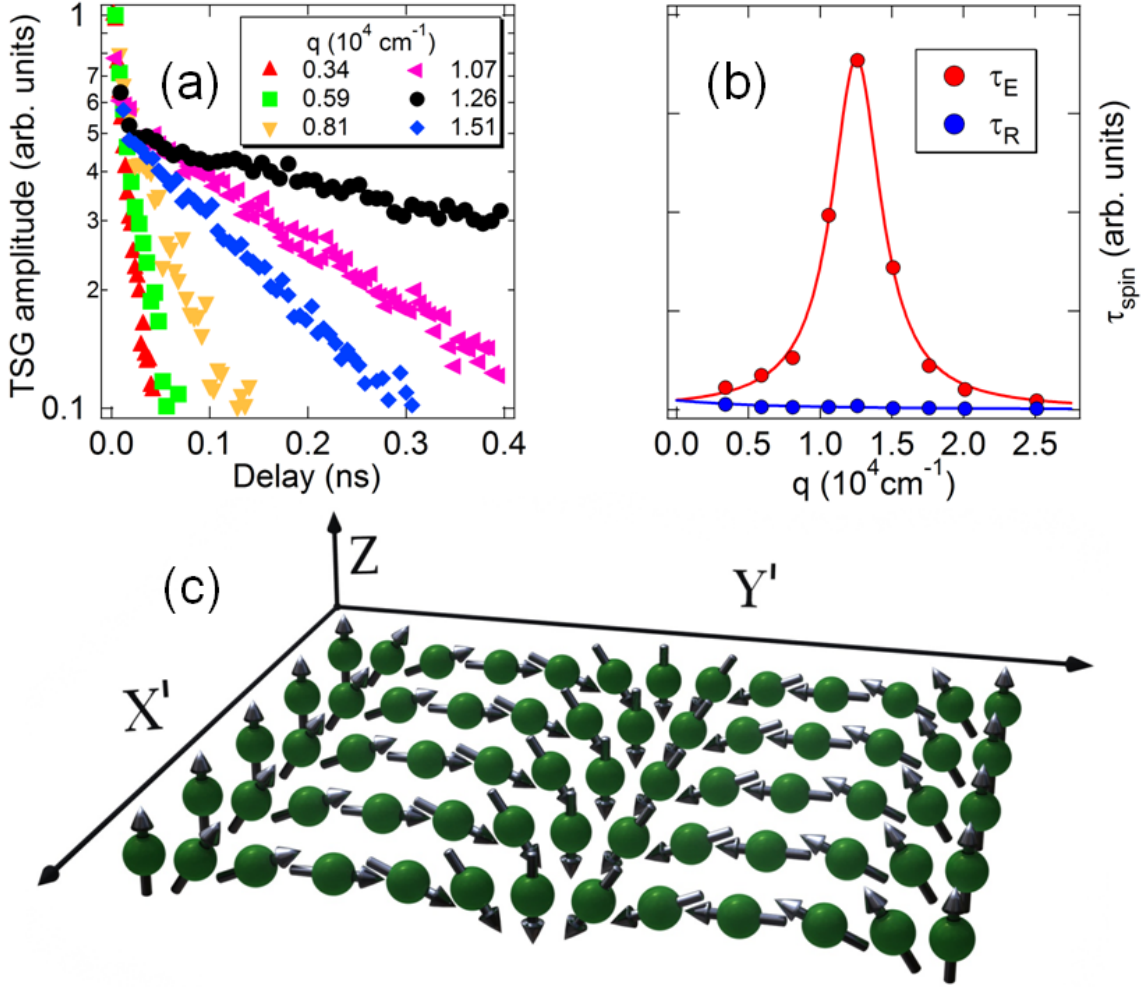

**Figure 1 | Bi-exponential decay of transient spin-gratings.** TSG decay curves at various wavevector **q** are shown in panel (a) for an asymmetrically doped 2DEG with a mixture of Rashba and Dresselhaus spin-orbit coupling. Panel (b) shows the lifetimes for the SO-enhanced ($\tau_E$) and reduced ($\tau_R$) helix modes extracted from double exponential fits to the data of (a). The solid lines are a theoretical fit, as described in the text, using a single set of SO parameters for both helix modes. Error bars (s.d.) are the size of the data points. (c) is an illustration of a helical spin wave, which is one of the normal modes of the SO coupled 2DEG. In this picture, z is the growth direction (001), and the axes x' and y' refer to the $[11]$ and $[1\bar{1}]$ directions in the plane of the 2DEG. The green spheres represent electrons whose spin directions are given by the arrows.

The salient features of Figs. 1a and 1b were, in fact, predicted by recent quantitative theories of spin propagation in a 2DEG in the presence of SO coupling. Froltsov[11] and Burkov, Nunez, and MacDonald[12] studied the effects on spin propagation of the Rashba interaction term in the Hamiltonian, $H_R = \alpha \cdot \hbar v_F (k_y \sigma_x - k_x \sigma_y)$, where $v_F$ is the Fermi velocity, $\hbar k$ is the electron momentum, and $\sigma_{x,y}$ are Pauli matrices. $H_R$ corresponds to an in-plane, k-dependent magnetic field, $\mathbf{b_R} = \alpha \hbar v_F (k_y \hat{\mathbf{x}} - k_x \hat{\mathbf{y}})$, that leads to precession of the electron's spin. It was found that, in the presence of $\mathbf{b_R}$, the normal modes of the system are helical waves of spin polarization, in which the spin direction rotates in the plane normal to the 2DEG and parallel to the wavevector, $\mathbf{q}$ (see Fig. 1c). For each $\mathbf{q}$ there are two helical modes with opposite sense of rotation. The lifetime of the mode whose sense of rotation matches the precession of the electron's spin is enhanced, while the other is reduced. A striking prediction is that, for a range of $\mathbf{q}$, the SO-enhanced lifetime will exceed that of a uniform ($q=0$) spin polarization. This stands in contrast to ordinary diffusion, where the decay rate for spin excitations goes as $q^2$, and the spin lifetime is always greatest at $q=0$. (The same conclusions are reached when the linear Dresselhaus term, $H_D = \beta_1 \cdot \hbar v_F (k_x \sigma_x - k_y \sigma_y)$, is assumed to be the only SO interaction). Weber et al.[13] provided experimental support for these predictions, observing a maximum spin lifetime at nonzero $q$ in nominally symmetric GaAs QW's, where the $H_D$ dominates the SO Hamiltonian.

Recently, this theory has been extended to predict the lifetimes of helix modes in the presence of both $H_R$ and $H_D$.[2] In particular, it was predicted as the two couplings approach equal strength, the SO-enhanced mode evolves to the PSH, i.e., the lifetime tends to infinity for $\mathbf{q}=\mathbf{q}_{PSH}$. As discussed above, the stability of the PSH is a manifestation of restored SU(2) symmetry at this point in parameter space. Although conservation of the x-component (see Fig. 1c) of spin, or U(1) symmetry, had been noted previously,[14] SU(2) symmetry implies conservation of the amplitude and phase of the PSH as well. The theory also predicts quantitatively how the persistence of the helix degrades with detuning from the SU(2) point, either by variation of $q$ or the $\alpha/\beta_1$ ratio. Stanescu and Galitski (SG)[15] extended the theory further by including the SU(2)-breaking effects of the $k^3$-Dresselhaus coupling,

$$H_{cD} = -4\beta_3 \cdot \frac{\hbar v_F}{k_F^2}\left(k_x k_y^2 \sigma_x - k_y k_x^2 \sigma_y\right)$$

which is always present at some level because of the nonzero width of the QW (vide infra).

The predictions of helical spin modes described above are clearly evident in the TSG results shown in Fig. 1. The initial condition created by the two pump pulses – a sinusoidal wave of $S_z$ at $t = 0$ – is equivalent to two equal amplitude $S_y$-$S_z$ helices of opposite pitch. Each of these normal modes then decays independently with its own characteristic decay rate, corresponding to the SO-enhanced and reduced helix lifetimes ($\tau_E$ and $\tau_R$ respectively). The reduced lifetimes shown in Fig. 1b peak at $q=0$, while the enhanced lifetime is greatest at finite $\mathbf{q}$. The solid lines are a fit to the SG theory using a single set of SO parameters for both the enhanced and reduced helix modes.

The fact that the dispersion of both branches is accurately fit by a single SO parameter set suggests that spin helices are indeed the normal modes of our SO-coupled 2DEGs. The theoretical fits provide us with values for $\alpha$, $\beta_1$, $\beta_3$ and $D_S$ (the spin diffusion coefficient), which we then use to guide us in engineering QWs with the longest spin helix lifetimes. To tune the SO Hamiltonian, we have designed a series of QW samples with varying doping asymmetry and well width.

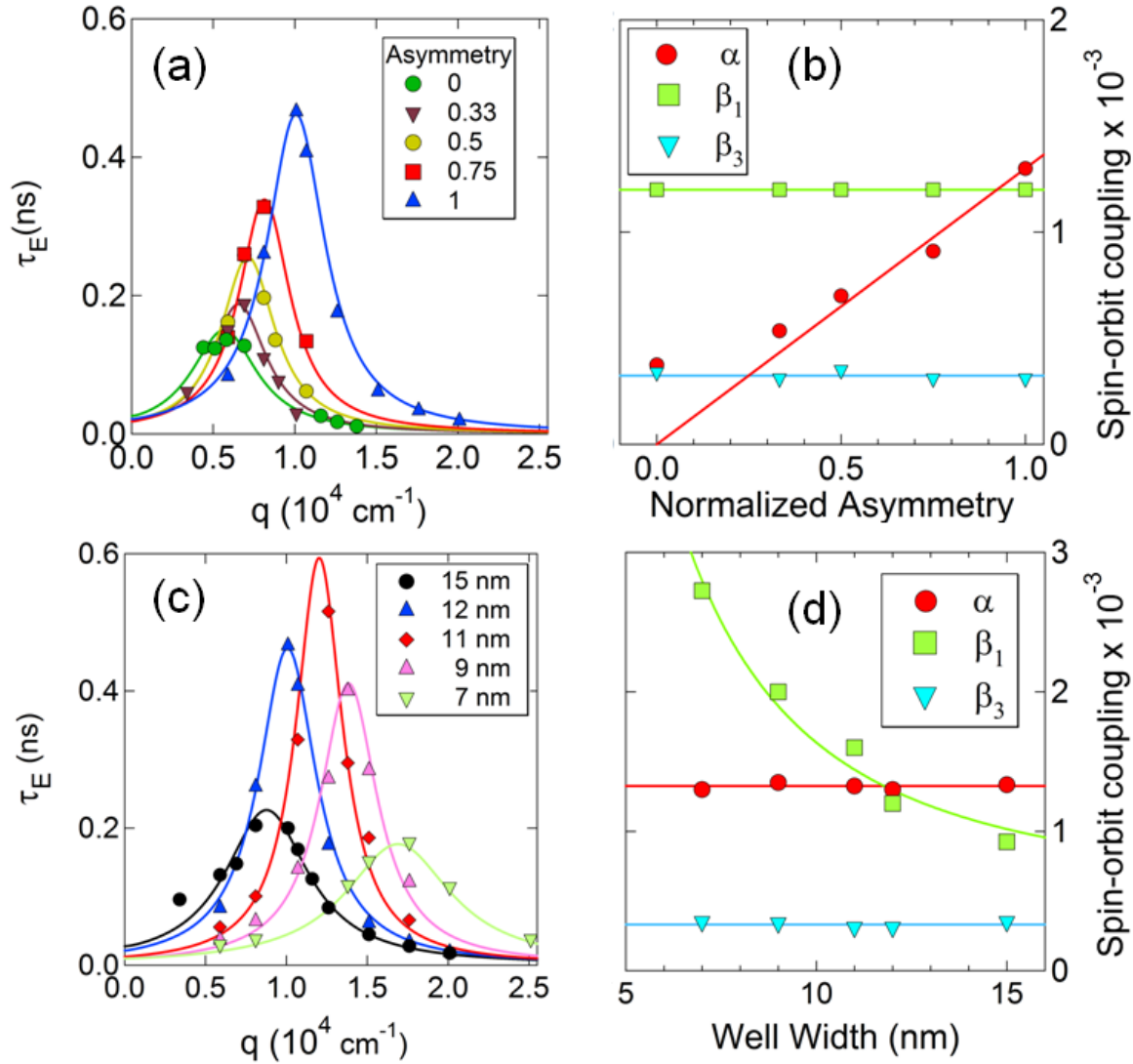

**Figure 2 | Rashba and linear Dresselhaus tuning.** Lifetimes of the enhanced helix mode are shown for samples with varying degrees of doping asymmetry (a) and well width (c). The normalized asymmetry is the difference in the concentration of dopants on either side of the well, divided by the total dopant concentration. The solid lines are fits to SG theory as described in the text. The SO parameters from these fits are summarized in (b) and (d). As in Ref. 15, the SO coupling strengths are expressed as dimensionless quantities by normalizing to the Fermi velocity.

Figure 2 summarizes the SO tuning results. These measurements were performed at T=75K, where the enhanced-mode lifetimes are greatest. (See discussion of *T* dependence below). To tune α, we varied the relative concentration of remote dopants on the two sides of the 2DEG while keeping the total carrier concentration fixed. Spin lifetimes are plotted vs. *q* in Fig. 2a, for a set of 12 nm wide QW's with varying amounts of doping asymmetry. The degree of lifetime enhancement grows monotonically with increasing dopant asymmetry. Fig. 2b shows the SO parameters extracted from comparison of the dispersion curves to the SG theory, for each of the samples. The parameters α, $\beta_1$, and $\beta_3$ are plotted as a function of normalized asymmetry, defined as the difference in the concentration of dopant ions on either side of the well, divided by the total dopant concentration. The variation of α is well approximated by a straight line that extrapolates to zero coupling as the asymmetry parameter goes to zero. The data for the nominally symmetric sample display a residual Rashba coupling, which we attribute to the inherent asymmetry in the growth of the QWs.[13,16,17,18] Although only $\tau_E$ is shown in Fig. 2a, the fast decay rates are accurately described by the same set of parameters.

We tuned $\beta_1$ by varying the width, *d*, of the QW's, with the normalized asymmetry fixed at unity. Experimentally determined spin lifetimes and theoretical fits are plotted in Fig. 2c for *d* ranging from 7-15 nm. Unlike the case of Rashba tuning, the amplitude and inverse width of the dispersion peak (Fig. 2c) show a clear maximum, suggesting that we have tuned through α = $\beta_1$ in this sample set. The curves generated from the SG theory fit the data very well with both α and $\beta_3$ remaining essentially constant, suggesting that we have indeed tuned the Dresselhaus interaction independently. The corresponding SO parameters, plotted in Fig. 2d, indicate that the crossing of α and $\beta_1$ occurs at a well width near 12 nm, which is consistent with the asymmetry series. The 11 nm sample actually has a longer lifetime, in agreement with SG theory which predicts the peak lifetime to occur near α = ($\beta_1-\beta_3$) rather than α = $\beta_1$.

As a check on the modeling of our TSG data, we compare the experimental values of α, $\beta_1$ and $\beta_3$ with band structure calculations. The Rashba coupling strength is predicted to obey $\alpha = r_{41}^{6c6c} e \langle E_z \rangle / \hbar v_F$, where $\langle E_z \rangle$ is the average electric field in the well, and $r_{41}^{6c6c}$ is an intrinsic proportionality factor. In **k**·**p** perturbation theory, the proportionality factor is found to be 5.206 Å$^2$ (Ref 19) for GaAs. To compare with theory, we assume that the electrons in the well experience the δ-layers as an infinite sheet of positive charge. We estimate the field strength to be $5.4 \times 10^6$ V/m for a normalized asymmetry of 1. From the corresponding value of *α*, we find $r_{41}^{6c6c} = 6.7$ Å$^2$, in good agreement with the perturbation theory result.

The Dresselhaus couplings, $\beta_1$ and $\beta_3$, are both proportional to a single intrinsic parameter, $\gamma$. While the commonly cited $\mathbf{k} \cdot \mathbf{p}$ value is $\gamma_{\mathbf{k}\bullet\mathbf{p}} = 27.58$ eV Å$^3$ (ref 19) for GaAs, recent calculations using the *ab initio* GW approach[20] suggest that this is an overestimate, and in fact $\gamma_{GW} = 6.5$ eVÅ$^3$. Most of the experimentally determined values fall between $\gamma_{\mathbf{k}\bullet\mathbf{p}}$ and $\gamma_{GW}$.[21] From the values of $\beta_1$ as a function of well width, determined by TSG spectroscopy and SG analysis, we find $\gamma = 5.0$ eVÅ$^3$, in excellent agreement with GW theory. The ratio $\beta_3/\beta_1$ is given theoretically as $k_F^2/4\langle k_z^2 \rangle$, *i.e.*, the ratio of the electron momentum parallel and perpendicular to the conducting plane. Estimating $k_z = \pi/d$, the expected ratio is 0.16, consistent with our experimental value of 0.2. The agreement between the predicted and experimental value for $\beta_3/\beta_1$ is evidence that the cubic Dresselhaus interaction limits the PSH lifetime at low temperature.

The temperature dependence of the spin-helix lifetimes further tests our understanding of 2DEG spin physics, and also is relevant to potential spintronics applications. The *T*-dependence of the lifetime of both modes, for the sample closest to the SU(2) point, is plotted on a log scale in Fig. 3a. The lifetime of the SO-enhanced mode increases with decreasing *T* to ~ 50 K, then drops rapidly with further lowering of *T*. The lifetime of the fast mode, on the other hand, decreases monotonically with decreasing *T*.

We cannot solely rely on the spin dynamics theories described previously to explain the observed *T*-dependence, as they consider only the *T* = 0 limit. However, these theories do indicate an important first step in the analysis. For a given set of SO parameters the spin-helix lifetimes for both senses of rotation, and for all *q*, are predicted to scale as $D_s^{-1}$. Because $D_s$ is known to depend strongly on *T*,[23] this scaling provides at least one well-understood mechanism for temperature-dependent lifetimes. However, the fact that the enhanced and reduced lifetime modes display very different *T*-dependence indicates immediately that scaling by $D_s(T)$ cannot fully account for the effect of temperature.

In order to focus on the *T*-dependent effects beyond scaling by $D_s$, we consider the dimensionless parameter, $\eta \equiv D_s q_{PSH}^2 \tau_{PSH}$, rather than $\tau_{PSH}$ itself (here the subscript PSH refers to quantities at the PSH wavevector). $\eta$ is the ratio of the measured PSH lifetime to what it would be in the absence of SO coupling (see methods), i.e., a direct measure of the lifetime enhancement as a result of proximity to the SU(2) point. Fig. 3b is a log scale plot of $\eta(T)$ for both helix modes. For the rapidly decaying mode $\eta_R$ is essentially independent of *T*, indicating that the temperature dependence of $\tau_R$ is entirely accounted for by $D_s(T)$. (The rapid increase of $D_s$ below 75 K is the result of the quenching of the spin-Coulomb drag effect.[22,23]) In contrast, $\eta_E$ for the SO-enhanced helix is strongly $T$-dependent. In this case normalization with respect to $D_s(T)$ converts the peak in $\tau_E$ near 75 K to a plateau at low $T$, demonstrating that the PSH lifetime enhancement is in reality a monotonically decreasing function of increasing $T$. The enhancement factor is approximately 100 at low $T$, and decays to towards unity (no SO enhancement) at high temperature. Above 50 K, the enhancement factor obeys the power law $\eta(T) \propto T^{-2.2}$. In Fig. 3c we plot the spin-lifetime dispersion curves for several values of $T$, illustrating the damping of the entire PSH resonance with increasing temperature. Although weakened, the PSH remains observable through room temperature. The existence of the PSH at temperatures far greater than the SO spin-splitting energy, which is only ~ 1K, supports the idea that the lifetime enhancement is symmetry driven.

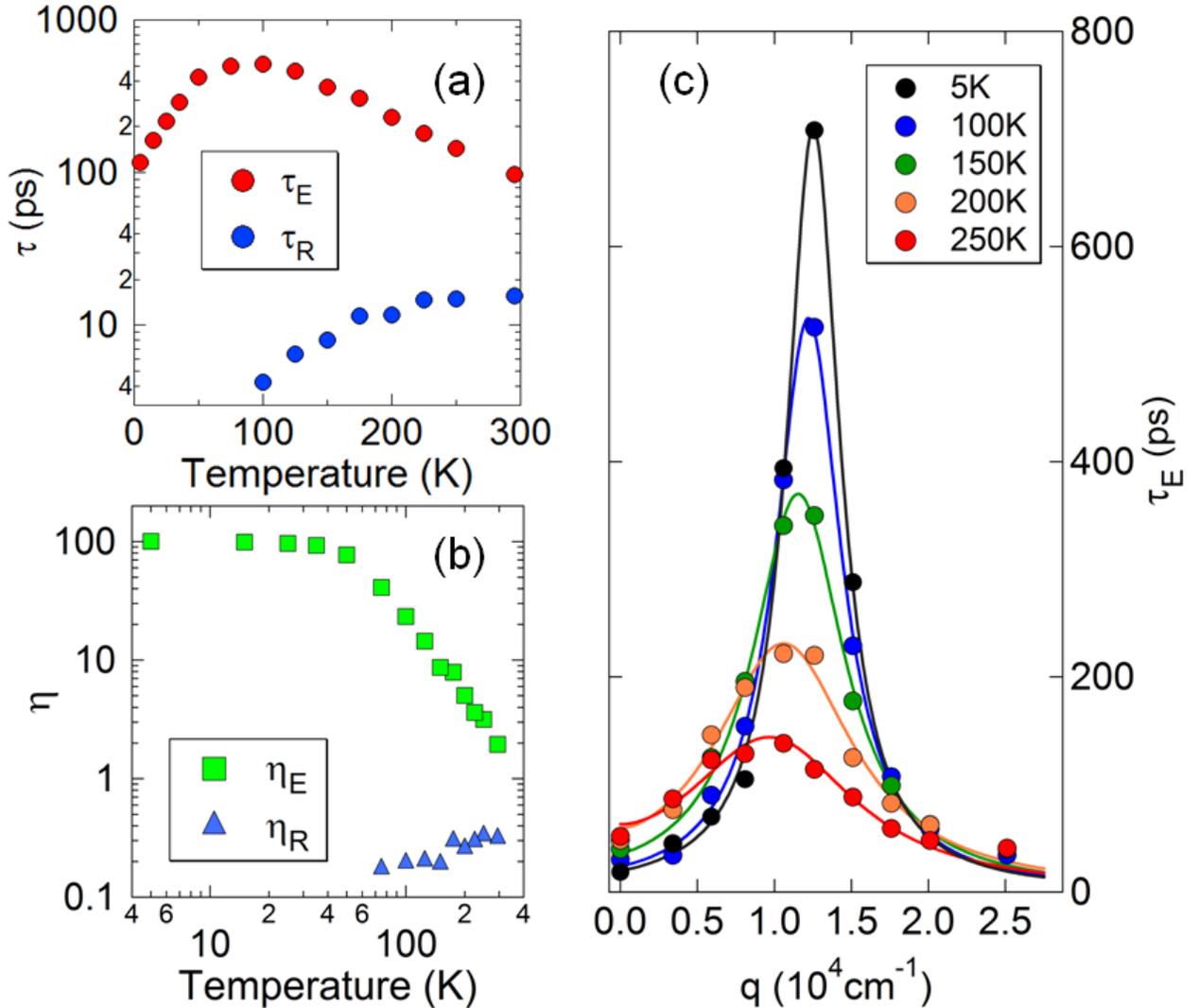

**Figure 3 | Temperature dependence of the persistent spin helix.** Temperature dependence of lifetimes for both helix modes at $q_{PSH}$ are shown in (a) for the 11 nm asymmetrically doped sample, which is the closest to the SU(2) point. Panel (b) shows the temperature dependence of the dimensionless lifetime-enhancement factor $\eta \equiv D_s q_{PSH}^2 \tau_{PSH}$ for both helix modes. (c) shows the temperature dependence of the PSH dispersion curves for a

similar sample with slightly reduced mobility. The reduced mobility suppresses the the drop in *Ds* and τ by avoiding the ballistic crossover. Fits to SG theory (solid lines) are also shown.

Why the PSH stability decreases strongly with *T* remains an open question. Within a non-interacting (or single-particle) picture, the cubic Dresselhaus term is the only SU(2)-breaking interaction. While the full theory is yet to be worked out, it is straightforward to show that when $\alpha = \beta_1$ the effect of small, but nonzero, $\beta_3$ is to reduce the helix lifetime by the factor $1/[1+(T/T_F)^2]$. This effect occurs because the cubic Dresselhaus interaction breaks the linear proportionality of the spin precession rate to velocity, making the PSH lifetime sensitive to thermal broadening of the velocity distribution. Although this effect gives the observed power law, it predicts, incorrectly, that the PSH is stable up to the Fermi temperature, which is ~400 K for our QW's. In considering effects beyond the single-particle picture, the approximate $T^{-2}$ scaling suggests a connection to electron-electron scattering. As mentioned previously, if SU(2) symmetry is exact, the PSH is not sensitive to electron-electron scattering.[2] It remains to be seen whether many body interactions can affect the PSH lifetime when SU(2) is weakly broken by the cubic Dresselhaus term, disorder in local SO couplings,[24] or spin-dependent scattering mechanisms.

Finally, we note that a PSH lifetime enhancement of 100 is not a fundamental limit. When controlled by the cubic Dresselhaus term, the lifetime enhancement is proportional to $(\beta_1/\beta_3)^2$. Gated structures in which electron density and electric field are tuned independently will enable this ratio to be increased, while maintaining $\alpha = \beta_1$. Because of the extreme sensitivity of the PSH lifetime to $\beta_3$, we believe that large lifetime increases are achievable. Increased stability of the PSH creates possibilities for new experiments on spin transport, such as measurement of intrinsic spin-Hall effect, charge transport dynamics in the presence of strong spatial variation of spin polarization, and demonstration of efficient spin-transistors.

## Methods

### Quantum well structures

The GaAs/ Al$_{0.3}$Ga$_{0.7}$As samples were grown in the (001) direction by molecular beam epitaxy, and consist of ten quantum wells separated by 48 nm barriers. The Si donors were deposited in eight single atomic layers in the centre 14nm of each barrier to maximize their distance from the 2DEG. The donor concentration in alternating barriers was adjusted as described in the text in order to tune α, while the total target carrier concentration in the wells was held fixed at n ≈ 6.5x10$^{11}$ cm$^{-2}$. The electron mobility typically reached µ ≈ 3 x 10$^5$ cm$^2$/Vs at low temperature. All samples were mounted on c-axis cut Sapphire discs, and the GaAs substrates were chemically etched to allow for spin grating measurements in the transmission geometry.

### Transient spin gratings

Transient spin gratings are generated by the optical interference of two cross-polarized pulses from a single modelocked Ti:Sapphire laser (80 MHz, 100 fs), focused non-collinearly onto the 2DEG. The time evolution of the spin excitation is monitored by time delayed probe pulses, which see the modulation of the 2DEG magnetization as a diffraction grating because of the Kerr effect. The diffracted probe is mixed with a redundant probe beam so that the amplitude and phase of the spin grating can be measured using a heterodyne detection scheme.[9]

### Determination of spin diffusion coefficient

We determine the spin diffusion coefficient, $D_s$, through measurement and analysis of the spin relaxation rate at $q = 0$. We have verified that $D_s$ obtained from the $q = 0$ data is consistent with that obtained from analysis of the full q dependence of the spin lifetimes. The decay of spin polarization at zero wavevector, $S_z(q=0,t)$, was measured by a standard time-resolved Faraday rotation technique.[10] As $T$ is reduced from room temperature, $S_z(q=0,t)$ crosses over from single exponential decay to decaying oscillations (at about 50 K in our QW's). The crossover occurs when the electron mean-free time becomes comparable to the period of precession in the SO effective fields. In the high-$T$ regime, we determine $D_s$ from the DP formula applicable when $\alpha=\beta_1$, which is $1/\tau_s = D_s q_{PSH}^2$. To determine $D_s$ through the crossover regime we use the phenomenological formula,

$$S_z(q=0,t) \propto \int_0^{\frac{\pi}{2}} \exp\left[-\left(\frac{\Omega\tau\cos\phi}{1-i\Omega\tau\cos\phi}\right)\Omega t\right] d\phi$$

where $\tau$ is the mean-free time, and $\Omega\cos\phi$ is the precession frequency as function of angle $\phi$ on the Fermi circle. This expression interpolates between the exact result for the $\Omega\tau \to 0$ limit, which is the zero-order Bessel function, and the non-oscillatory decay in the $\Omega\tau \gg 1$ limit.

**Acknowledgements:** We thank Keith Bruns for creating the PSH diagram of Fig 1c. S.M. thanks J. H. English and A. W. Jackson for MBE work, and J. Stephens and A. C. Gossard for helpful discussions. This work was supported by the US Department of Energy, Office of Science, under Contract DE-AC02-05CH11231, the National Science Foundation, and the Office of Naval Research.

**Author Information:** Correspondence and requests for materials should be addressed to JDK (JDKoralek@lbl.gov)